\documentclass[nofootinbib,floatfix,superscriptaddress,prd,twocolumn]{revtex4-1}
\usepackage{graphicx}
\usepackage{epsfig}
\usepackage{bm}
\usepackage{comment}
\pdfoutput=1
\usepackage[T1]{fontenc}
\usepackage[latin9]{inputenc}
\usepackage{amssymb}
\usepackage{float}
\usepackage{amsmath}
\usepackage{dcolumn}
\usepackage{cancel}

\usepackage[dvipsnames]{xcolor}

\usepackage[colorlinks]{hyperref}
\hypersetup{
    breaklinks=true,
    pdfstartview={FitH},    
    colorlinks=true,       
    linkcolor=blue,          
    citecolor=red,        
    filecolor=magenta,      
    urlcolor=magenta,           
    anchorcolor=green,      
    linktocpage=true
}

\newcommand{\be}{\begin{equation}}
\newcommand{\ee}{\end{equation}}

\begin{document}

\title{New modified cosmology from a new   generalized entropy}

\author{G.~G.~Luciano}
\email{giuseppegaetano.luciano@udl.cat}
\affiliation{Department of Chemistry, Physics and Environmental and Soil Sciences, Escola Politecnica Superior, Universidad de Lleida, Av. Jaume
II, 69, 25001 Lleida, Spain}
\author{E. N. Saridakis}
\email{msaridak@noa.gr}
 \affiliation{National Observatory of Athens, Lofos Nymfon, 11852 Athens, 
Greece}
\affiliation{CAS Key Laboratory for Researches in Galaxies and Cosmology, 
Department of Astronomy, University of Science and Technology of China, Hefei, 
Anhui 230026, P.R. China}
 \affiliation{Departamento de Matem\'{a}ticas, Universidad Cat\'{o}lica del 
Norte, 
Avda.
Angamos 0610, Casilla 1280 Antofagasta, Chile}

\begin{abstract}
We develop new modified  cosmological scenarios by applying the first law of 
thermodynamics at the Universe horizon, utilizing a new entropic 
functional that generalizes the standard Boltzmann-Gibbs-Shannon entropy. 
In particular, starting  from the general theory of entropy
in terms of the probability distribution over the accessible microstates, and 
by imposing violation of the separability requirement and thus  considering a 
generalized microstate scaling, we result to a generalized entropy 
expression, which applied   in systems with boundaries yields a
generalized  holographic-like area-law scaling   with two exponents.
Hence, incorporating it within the gravity-thermodynamics framework    we  
result to a modified cosmological 
scenario  with additional terms   which eventually give rise to  an 
effective dark energy sector. We extract analytical expressions for the dark 
energy density  and equation-of-state parameters, and we show that the 
Universe experiences the usual thermal history, with 
the sequence of matter and dark-energy eras. Additionally, depending on 
the values of the entropic exponents, the dark energy can 
be quintessence-like, phantom-like or experience the phantom-divide crossing 
during its evolution, ultimately stabilizing at the cosmological constant value 
  in the asymptotic far future, a behavior richer than other entropic modified 
cosmologies.
 
 \end{abstract}

 \maketitle

\section{Introduction}
 
There is   a well-known conjecture that gravity and thermodynamics are 
related, or equivalently that one can construct the thermodynamics of spacetime 
itself \cite{Jacobson:1995ab,Padmanabhan:2003gd,Padmanabhan:2009vy}.
In this framework one considers the universe as a thermodynamical system, and by 
applying the first law of thermodynamics, interpreted in terms of energy flux 
through local Rindler horizons, in its boundary, namely  the apparent horizon
\cite{Frolov:2002va,Cai:2005ra,Akbar:2006kj}, one can extract the   Friedmann 
equations. The basic application of gravity-thermodynamics conjecture has been 
performed in the framework of general relativity, resulting in the standard 
Friedmann equations of $\Lambda$CDM cosmology.

However, one can extend the whole analysis in two ways. The first is to 
maintain standard thermodynamics, but 
 consider various modified theories of gravity and thus their corresponding 
modified entropy relations \cite{Wang:2005bi, Akbar:2006er,Cai:2006pa, 
Sheykhi:2007zp, Wu:2007se,MohseniSadjadi:2007zq,Cai:2007bh,  
Akbar:2008vz, Cai:2008ys, Sheykhi:2008qs, Wu:2008ir, Bamba:2009ay, 
Jamil:2009eb,Wang:2009zv,Cai:2009qf, Bamba:2009id, Cai:2009ph, 
Jamil:2010di, 
Bamba:2010kf,Sadjadi:2010kp, Hendi:2010xr, Bamba:2011pz, Karami:2012fu, 
Bamba:2012rv,Sharif:2012zzd,Karami:2012hq,Salako:2013gka,Cao:2013xy,
Sharif:2013tha,Gim:2014nba, Fan:2014ala,Karami:2014tsa, Momeni:2015fyt, 
Chakraborty:2015wma, Zubair:2015yma, Zubair:2016uhx, Bahamonde:2016cul, 
Zubair:2016bpi, Rudra:2020rhs,Pourhassan:2020jfu,Jawad:2020wlg, 
Tsilioukas:2023tdw, Rao:2024ncj, Tsilioukas:2024seh, Petronikolou:2025mlm, 
Lymperis:2025vup}. The second is  to maintain general 
relativity but use various extended entropies that arise from extended 
thermodynamics \cite{Lymperis:2018iuz, Moradpour:2016rcy,Moradpour:2017ycq, 
Moradpour:2018ivi,MohseniSadjadi:2010nu, Bamba:2018zil,Nojiri:2019skr,  
Nojiri:2019itp,  Hernandez-Almada:2021rjs, 
 Asghari:2021bqa,  Nojiri:2021jxf,  Saridakis:2020lrg, 
Giardino:2020myz, Barrow:2020kug, Leon:2021wyx,
Lymperis:2021qty, Ghoshal:2021ief, Jizba:2022bfz,Keskin:2021zct,  
Odintsov:2022qnn,   Luciano:2022ely,Dheepika:2022sio,  Luciano:2022knb,   
Nojiri:2022nmu,  Luciano:2022pzg,  Basilakos:2023kvk,
Lambiase:2023ryq, Mohammadi:2023lss, Salehi:2023gzv, 
Jizba:2023fkp, Luciano:2023roh,  Jizba:2024klq, Petronikolou:2024zcj, 
 Mondal:2024wno, Asghari:2024sbu, Karabat:2024trf}. Note that although in both 
cases one obtains  modified 
 Friedmann equations, i.e. modified cosmologies, in the latter case the 
modified cosmologies are obtained without the need to modify gravity, which is 
still general relativity, and that is why it has attracted the interest of the 
community.

There are many   reasons that lead to the need of extending standard 
thermodynamics and in particular standard entropy. In standard thermodynamics, 
the Boltzmann-Gibbs-Shannon  (BGS) entropy serves as a fundamental concept in 
the statistical description of numerous equilibrium systems, and it is defined 
as  
\begin{equation}  
\label{SBG}  
S_{\text{BGS}} = -\kappa \sum_{i=1}^W p_i \ln p_i\,,  
\end{equation}  
where \( p_i \) denotes the probability distribution  over the accessible 
microstates \( W \). This entropy has been instrumental in advancing our 
understanding of the microscopic foundations of various macroscopic 
phenomena (note that in the context of holographic thermodynamics  the factor 
\( \kappa \) in Eq. \eqref{SBG} is typically chosen as the Boltzmann constant 
\( k_B \)) \cite{goldstein2020gibbs}.  

Despite its broad applicability, the BGS  entropy encounters limitations 
when applied to complex systems, particularly those exhibiting non-equilibrium 
dynamics, long-range interactions, strong correlations or memory effects. In 
such cases, the conventional formalism may fail to encapsulate the full 
statistical complexity, leading to discrepancies between theoretical 
predictions 
and experimental observations. Consequently, there has been increasing interest 
in extending the definition \eqref{SBG} to encompass systems characterized by 
``extended'' statistical behavior.  

A crucial consideration in generalizing the BGS entropy is that, in  a 
thermodynamic context, it links the number of accessible states to an 
 extensive  property. To maintain the possibility of describing complex 
systems using thermodynamic formalism, it is imperative to identify an 
appropriate entropic form adapted to the specific statistical features of the 
system under study. 

In a broad sense, generalized entropies can be expressed  in the form 
\cite{hanel2011comprehensive}  
\begin{equation}  
\label{Sf}  
S_f[p] = \sum_i f(p_i)\,.  
\end{equation}
From an information-theoretic perspective, if one   imposes the    four 
Khinchin axioms (K1-K4), namely:
\begin{itemize}
    \item K1: \textbf{Continuity} - The condition that $S$ varies continuously 
with $p$ necessitates that $f$ is a continuous function.
    \item K2: \textbf{Maximality} - The requirement that $S$ is maximal for a 
uniform distribution (i.e.  $p_i=1/W$) entails that $f$ is a concave function.
    \item K3: \textbf{Expandability} - Adding an event with zero probability 
does not affect the entropy. This implies that $f(0)=0$.
    \item K4: \textbf{Separability} - The entropy of a joined system $A \cup B$ 
can be expressed as $S(A \cup B) = S(A) + 
S(B\hspace{-0.8mm}\mid\hspace{-0.8mm}A)$, where 
$S(B\hspace{-0.8mm}\mid\hspace{-0.8mm}A)=\sum_i p_i^A \hspace{1mm} 
S(B\hspace{-0.8mm}\mid\hspace{-0.8mm}A=a_i)$. 
\end{itemize}
then one   
uniquely determines $f$ as the BGS entropy in 
Eq. \eqref{SBG} \cite{shannon1948claude,Khinchin1957}. 

However, while axioms K1-K3 are generally valid, K4 is violated for most 
interacting systems. In this case, introducing a generalized $f[p]$ form may 
provide  a valid approach for describing interacting statistical systems within 
the framework of thermodynamic formalism, and ensure the extensivity of 
entropy. 

In this work we present a new extended entropy that violates requirement K4. In 
particular, we propose an     entropy expression $f[p]$ being a sum of two 
terms, which corresponds to a generalized microstate scaling, which eventually 
give rise to a     black-hole entropy expression being the sum of two 
area terms. Hence, when applied in the framework of gravity-thermodynamics 
conjecture, this 
extended entropy gives rise to modified Friedmann equations, and thus to 
modified cosmology. 

The remainder of the work is organized as follows: In Section \ref{Themodel}, 
we construct the new  extended entropy in detail, and we apply it in systems 
with boundaries. Then in Section \ref{ModCosm} 
we present the derivation of the extended cosmology arsing from the 
gravity-thermodynamics conjecture with the above    modified horizon 
entropy. In Section \ref{Imp} we investigate the cosmological implications of 
the scenario at hand. Finally,  Section \ref{CandO} is dedicated to 
conclusions and outlook.

\section{A new     generalized entropy }
\label{Themodel}
 
In this section we construct   an extended entropy expression by choosing 
suitably the  $f[p]$ form in (\ref{Sf}). We start by recalling that 
  (large) statistical systems have been 
classified according to the asymptotic properties of their associated 
generalized entropies\footnote{By ``asymptotic'', it is meant   that the 
total number of available states is large enough, namely  $W \gg 1$. Under this 
assumption, all relevant entropic information about the system is contained in 
the behavior of $f(x)$ near zero, specifically in the region $x \sim W^{-1}$.} 
\cite{hanel2011comprehensive}. Specifically, it has been shown that all 
entropies of the form \eqref{Sf} that satisfy requirements K1-K3 can be 
encompassed by the 
two-parameter family 
\begin{equation}
 S_{c,d} \propto \sum_i \Gamma\left(d+1, 1-c\log 
p_i\right),
\label{entrclass}
\end{equation}
where $\Gamma$ represents the incomplete Gamma function, and the 
exponents $c$, $d$ characterize the scaling functions associated with each of 
the two asymptotic properties employed for the classification (see also 
\cite{PhysRevE}). Notable examples include generalized entropies that are 
widely 
used in the literature to study complex systems,
such as the $q$-Tsallis  entropies \cite{Tsallis:1987eu,Tsallis:2009}
\begin{equation}
 c=q<1, d=0\Longrightarrow 
S_{q<1}=\frac{1-\sum 
p_i^q}{q-1}
\end{equation}
 and   
 \begin{equation}
  c=1,d=0\,\Longrightarrow 
S_{q>1}=\frac{1-\sum p_i^q}{q-1},
 \end{equation}
  the
Kaniadakis entropy \cite{kaniadakis2001non,Kaniadakis:2002zz}) 
 \begin{equation}
  c=1-\kappa, d=0\Longrightarrow S_{0<\kappa\le 1}=-\sum_i 
p_i \frac{p_i^\kappa-p_i^{-\kappa}}{2\kappa},
 \end{equation}
and the Curado-Nobre entropy \cite{CURADO200494})
 \begin{equation}
  c=1,d=0\Longrightarrow S_{b>0}=\sum_i (1-e^{-bp_i})+e^{-b}-1.
 \end{equation}
One may see other  examples of   class (\ref{entrclass})  in 
\cite{hanel2011comprehensive}.

A special case of the $c-d$ entropy class  (\ref{entrclass})  is obtained   for 
$c=1, d=\delta>0$, and is the so-called Tsallis-Cirto (or $\delta$-Tsallis) entropy 
\cite{Tsallis:2009, Tsallis:2013}, which 
represents a generalized entropic functional of the form
\begin{equation}
\label{Tsallis}
    S_{\delta}\,=\,\eta_\delta \sum_i p_i\left(\log \frac{1}{p_i}\right)^\delta\,,
\end{equation}
where the positive constant $\eta_\delta$  (which is generally dependent on 
$\delta$)  reflects the units used for measuring entropy. The standard BGS 
statistics is straightforwardly recovered in the limit $\delta=1$.
For a uniform distribution, using the asymptotic equipartition property 
\cite{Cover2006}, $S_\delta$ can be equivalently  expressed in its 
microcanonical form as
\begin{equation}
\label{Tsallis2}
    S_\delta\,=\, \eta_\delta \left(\log W\right)^\delta\,.
\end{equation}

One of the contexts  where the $\delta$-Tsallis entropy is most commonly applied 
is in systems characterized by a sub-extensive scaling of microstates, such as 
quantum condensed-matter systems \cite{Srednicki:1993im,Eisert:2008ur}, black 
holes \cite{Bekenstein:1973ur, Bekenstein:1974ax, Hawking:1975vcx, 
Gibbons:1976ue}, and more generally gravitational systems such as
the entire Universe \cite{Bousso:2002ju, Cai:2005ra, Padmanabhan:2003gd}. In 
fact, unlike the BGS entropy, $S_\delta$ can be regarded as a valid 
thermodynamic entropy for such systems, provided that $\delta$ is appropriately 
selected. To understand this, consider, for example, the black hole entropy.
Within the standard BGS statistics based on Eq. \eqref{SBG}, this entropy  
exhibits the holographic-like area-law scaling,  namely 
\cite{tHooft:1993dmi,Susskind:1994vu}
\begin{equation}
\label{arealaw}
S_{BGS}\,\propto\,\log W\,\propto\, L^2\,,
\end{equation}
where $L$ represents the characteristic size of the system.  This implies that 
the total number $W$ of available microstates scales exponentially, i.e.
\begin{equation}
W \, = \, g(L)\, \xi^{L^2}\, , \quad \mbox{with} \;\;\; \xi >1 \, ,
\label{W}
\end{equation}
where $g(L)$ ``weakly'' depends on $L$, so that $\log\left(g(L)\right)$ tends to 
 zero more rapidly than any positive power of $L$ for very large size $L$. 
If black holes are to be considered as genuine $d = 3$ systems (which is 
reasonable given that the corresponding spacetime is $(3+1)$-dimensional), it is 
evident that the scaling in Eq. \eqref{W} prevents the BGS entropy from being 
regarded as a fully thermodynamic (i.e.  extensive) entropy.

On the other hand, as argued in \cite{Tsallis:2013},  $S_\delta$ may be 
considered as a proper thermodynamic entropy  as it
preserves the desired structure of thermodynamic  Legendre transforms and, 
additionally, it becomes extensive for suitable $\delta$ 
\cite{Tsallis:2013,Tsallis:2019giw}.
This can be easily seen by noting that  in the limit of large $L$ relations
\eqref{Tsallis2} and \eqref{W} entail the generalized entropy-area law 
\begin{equation}
\label{TsArLaw}
    S_\delta \,=\, \Gamma_\delta \left(L^2\right)^\delta\,,
\end{equation}
where $\Gamma_\delta$ is a  $\delta$-dependent constant, which for $\delta=1$ 
reduces to Hawking's conventional form $\Gamma_{\delta}=1/(4\ell_p^2)$ (in units 
of $k_B$), with $\ell_p$   the Planck length.
As we observe,   $\delta>1$ ($\delta<1$) implies that the number of quantum 
microstates describing a black hole in Tsallis  picture is higher (lower) than 
in the semi-classical situation.
Clearly, when the number of microstates scales according to relation 
\eqref{W}, the exponent $\delta$ should be assigned the value of $3/2$ in three 
spatial dimensions to ensure that the entropy remains an extensive thermodynamic 
quantity \cite{Tsallis:2013,Tsallis:2019giw}. 

At this point  let us mention that  a similar power-law extension of 
the Bekenstein-Hawking entropy was proposed by Barrow \cite{Barrow:2020tzx}, 
based on a deformed holographic scaling at the quantum-gravitational level. 
This extension is attributed to the presumed fractal (sphereflake-like) 
structure of a black hole  surface resulting from quantum fluctuations 
\cite{Jalalzadeh:2021gtq,Jalalzadeh:2022uhl} (see also 
\cite{Buoninfante:2020guu}). The Barrow entropy takes the form
\begin{equation}
    S_\Delta\,=\,\Gamma_\Delta \left(L^2\right)^{1+\frac{\Delta}{2}}\,,
    \label{DeltaB}
\end{equation}
where $\Delta$ represents the \emph{anomalous dimension}, quantifying the 
deviation of the scaling dimension (namely  $2 + \Delta$) from its classical 
value (i.e.  $2$).  From expression \eqref{DeltaB}, one can derive the modified 
scaling \cite{Jizba:2024klq}
\begin{equation}
    W_\Delta\,=\,g(L)\,\xi^{L^{2+\Delta}}\,, \quad \xi>1\,.
\end{equation}
Although the anomalous dimension  in the original Barrow model is constrained 
within the range $0 \leq \Delta \leq 1$ (where $ \Delta = 1 $ corresponds to the 
most complex, fractal-like structure and $ \Delta = 0 $ represents the simplest 
classical geometry), renormalization group analysis in quantum field theory also 
supports the possibility of negative values $\Delta$ in various systems 
\cite{Dagotto:1989gp}. Consequently, it is generally expected that  $\Delta $ 
lies within the range $(-1, 1]$.

Both the $\delta$-Tsallis  and Barrow entropies have been extensively utilized 
in the study of black hole thermodynamics and cosmology, providing enriched 
phenomenological models that extend beyond conventional descriptions (see  
e.g.  \cite{Tavayef:2018xwx,Saridakis:2018unr,     
DAgostino:2019wko,   Saridakis:2020zol,  
Srivastava:2020cyk, Abreu:2020wbz, Dabrowski:2020atl, 
  Anagnostopoulos:2020ctz, Nojiri:2022aof,  
Jusufi:2021fek, Luciano:2023fyr, Luciano:2023zrx, Luciano:2023wtx, 
 DAgostino:2024sgm,Luciano:2025hjn,Luciano:2025elo} and references therein). 

Now we have all the material to construct a new generalized entropy. 
Within the  axiomatic  formulation illustrated above, one may consider 
introducing a generalization of BGS entropy, namely an  $f[p]$ function in 
(\ref{Sf}) of the form:
\begin{eqnarray}
\nonumber
    S_{\delta,\epsilon}&=&\eta_\delta \sum_i  p_i \left(\log 
\frac{1}{p_i}\right)^\delta\,+\,\eta_\epsilon \sum_i p_i \left(\log 
\frac{1}{p_i}\right)^\epsilon\\[2mm]
    &=&\eta_\delta\left(\log W\right)^\delta\,+\,\eta_\epsilon\left(\log 
W\right)^\epsilon\,, 
\label{genentropyexpr}
\end{eqnarray}
for equiprobable distributions, where $\delta,\epsilon>0$. 
This expression can be equivalently understood as resulting from a generalized microstate scaling, given by
\begin{equation}
\label{microscaling}
    W_{\delta,\epsilon}\,=\, g(L)\, \xi^{L^{2\delta}}\,\tilde{\xi}^{L^{2\epsilon}}\,,\quad \xi, \tilde\xi>1\,.
\end{equation} 
In fact, by inserting $W_{\delta,\epsilon}$ into Eq. \eqref{arealaw}, we obtain 
the modified area-law entropy
\begin{eqnarray}
\nonumber
S_{\delta,\epsilon}&=&\gamma_\delta \left(L^2\right)^\delta\,+\,\gamma_\epsilon \left(L^2\right)^\epsilon\\[2mm]
&=&\gamma_\delta A^\delta\,+\, \gamma_\epsilon A^\epsilon\,, 
\label{Sde}
\end{eqnarray}
in the limit of large size,  where $A=L^2$ is the area of the boundary surface 
and $\gamma_\delta, \gamma_\epsilon$ are suitable positive constants with 
dimensions $[L^{-2\delta}]$ and $[L^{-2\epsilon}]$, respectively. 

We note that the generalized microstate scaling \eqref{microscaling} can be viewed as a natural multiscaling extension of standard holographic state growth, as often observed in complex systems exhibiting hierarchical correlations and multifractal structures (see, e.g., \cite{Fractal, hanel2011comprehensive}). 
In this sense, the parameters $\delta$ and $\epsilon$  quantitatively capture the intricate nature of the system.  
It is straightforward to verify that
 the Bekenstein-Hawking 
formula is retrieved in   three
limiting cases, namely
\begin{subequations}
\label{limits}
\begin{align}
      &\hspace{-4mm} 1.\quad\,\,\,\,\, \delta=1, \quad  
\gamma_{\delta}=1/(4\ell_p^2), \quad \gamma_\epsilon=0; \\[2mm]
      &\hspace{-4mm} 2.\quad\,\,\,\,\, \epsilon=1, \quad \gamma_{\delta}=0,  
\quad\gamma_\epsilon=1/(4\ell_p^2); \\[2mm]
      &\hspace{-4mm} 3. \quad\,\,\,\,\, \delta=\epsilon=1, \quad 
\gamma_\delta=\gamma_\epsilon=1/(8\ell_p^2).
\end{align}
\end{subequations}
On the other hand,  the generalized entropy 
$S_{\delta,\epsilon}$ recovers the $\delta$-Tsallis  and Barrow entropy  as a 
special case when $\delta = \epsilon$. 

Let us make here an important point. In the above construction we started from 
the fundamental entropy expression (\ref{Sf}) in terms of the probability 
distribution over the   accessible microstates, and by considering the 
generalized microstate scaling (\ref{microscaling}) we resulted to the 
generalized entropy expression (\ref{genentropyexpr}). And then, by applying it 
in systems with boundaries, we extracted the holographic-like area-law scaling   
\eqref{Sde}. In \cite{Nojiri:2022aof} a generalized area-law entropy similar to 
\eqref{Sde} was proposed, but in an artificial way, namely straightaway at the 
level of area laws. However, in the present 
work we do provide the microscopic foundations of our new generalized entropy 
expression in a consistent way, starting from the violation of the 
separability requirement K4 and its effect on the probability 
distribution.

The generalized entropy expression  \eqref{Sde} will serve as the foundation 
for our subsequent analysis. In particular, we will apply it in the 
gravity-thermodynamics framework, obtaining novel cosmological scenarios, 
phenomenologically richer than $\Lambda$CDM paradigm. 
Specifically, we will derive modified Friedmann equations that, in the general 
case, incorporate additional terms, while reducing to the conventional form 
when the generalized entropy recovers the standard Bekenstein-Hawking 
expression.

\section{Modified cosmology through modified horizon entropy}
\label{ModCosm}

The ``gravity-thermodynamics''  conjecture establishes a deep connection between 
gravitational phenomena and thermodynamic principles, suggesting that the 
Einstein field equations can be derived from thermodynamic concepts such as 
entropy, temperature and energy flow 
\cite{Jacobson:1995ab,Padmanabhan:2003gd,Padmanabhan:2009vy}. 
This conjecture  interprets gravitational systems as thermodynamic systems, with 
black hole thermodynamics serving as a key analogy. In particular, in 
cosmological contexts, it can be shown that the Friedmann equations can be 
expressed as the first law of thermodynamics when the Universe is considered a 
thermodynamic system bounded by the apparent horizon 
\cite{Frolov:2002va,Cai:2005ra,Akbar:2006kj}. Similarly, by reversing the 
procedure, the Friedmann equations can be derived by applying the first law of 
thermodynamics.  In the following, we first review the procedure in the case of 
standard general relativity and strandard thermodynamics, and then we will 
apply it in the case of our new generalized entropy  \eqref{Sde}.

\subsection{Friedmann equations from the first law of thermodynamics}

Let us  consider an expanding homogeneous and isotropic Universe in $(3+1)$ 
dimensions, described by the Friedmann-Robertson-Walker (FRW) metric (we adopt 
units where $\hbar=c=k_B=1$)
\be
\label{FRW}
ds^2\,=\,h_{bc}\hspace{0.2mm}dx^{b}dx^{c}+\tilde r^2\left(d\theta^2+\sin^2\theta\,d\phi^2\right),
\ee
where $\tilde r=a(t)r$ is the proper distance at cosmic time $x^0=t$, $x^1=r$ is the comoving radius and $h_{bc}=\mathrm{diag}[-1,a^2/(1-kr^2)]$ denotes the metric
of the $(1+1)$-dimensional subspace.  
The curvature $k=0,+1,-1$ corresponds to flat, close and open spatial geometry, 
respectively. The information about cosmic expansion is encoded in the scale 
factor $a(t)$, which governs the growth of distances over time. It relates to 
the redshift through
\begin{equation}
\label{reds}
    a\,=\,\frac{1}{1+z}\,,
\end{equation}
and its derivative is related to the  Hubble parameter through $H=\dot a/a$, 
(the overdot denotes time derivative).

Additionally,  we assume that the energy content of the Universe is in the form 
of a perfect matter fluid, described by the stress-energy tensor
\begin{equation}
T_{\mu\nu}\,=\,\left(\rho_m+p_m\right)u_\mu u_\nu+p_m g_{\mu\nu}\,,
\end{equation}
where $\rho_m$, $p_m$ and $u^{\mu}$ indicate  the energy density, pressure and 
four-velocity of the fluid, respectively. 
Since we aim to study the effects of generalized  entropy on the evolution of 
the Universe at present time, we can safely neglect the radiation component 
(although this can be included straightforwardly).
The total energy content  of the Universe satisfies the conservation equation $ 
\nabla_\mu T^{\mu\nu} = 0$, which leads to the continuity equation:
\begin{equation}
\label{cont}
   \dot{\rho}_m + 3H(\rho_m + p_m) = 0\,. 
\end{equation}

According to the gravity-thermodynamic conjecture,  the first law can be 
understood as describing the energy flux across local Rindler horizons. 
Specifically, it is applied to the Universe's horizon, treated as a 
thermodynamic system separated by a causality barrier 
\cite{Jacobson:1995ab,Padmanabhan:2003gd,Padmanabhan:2009vy}. Such horizon is 
typically regarded as the apparent horizon \begin{equation}
\label{ra}
    \tilde r_A\,=\,\frac{1}{\sqrt{H^2+\dfrac{k}{a^2}}}\,.
\end{equation}
In the following discussion  we will focus on the flat case ($k=0$) for 
simplicity. This assumption is supported by recent observational data from the 
Cosmic Microwave Background (CMB) and Baryon Acoustic Oscillations (BAO) 
\cite{Planck:2018vyg}. Thus, the relation \eqref{ra} takes the form $\tilde 
r_A=1/H$.

To apply the first law of thermodynamics   it is necessary to assign both 
entropy and temperature to the Universe horizon. These quantities are derived 
from black hole thermodynamics, where the black hole horizon is replaced by the 
cosmological apparent horizon. For black holes, it is known that the definition 
of temperature follows from that of surface gravity 
$\kappa=\frac{1}{2\sqrt{-h}}\partial_a\left(\sqrt{-h}\hspace{0.2mm}h^{ab}
\partial_b\tilde r\right)$, i.e. \cite{Akbar:2006kj}
\begin{equation}
    T_h\,=\,\frac{\kappa}{2\pi}\,=\,-\frac{1}{2\pi\tilde r_A} \left(1-\frac{\dot 
{\tilde r}_A}{2H\tilde r_A}\right).
    \label{temp}
\end{equation}
 At this point, it is important to emphasize that in the following analysis we 
adopt the equilibrium assumption, according to which the temperature of the 
horizon is taken to be equal to that of the Universe fluid. This assumption 
holds reasonably well for the late-time Universe 
\cite{Padmanabhan:2009vy,Frolov:2002va,Cai:2005ra,Akbar:2006kj,Jamil:2010di,
Izquierdo:2005ku}.

The black hole entropy, which depends on the underlying  gravitational theory, 
in the case of General Relativity (GR)  is given by the standard 
Bekenstein-Hawking relation $ S = A/(4G)$ (see   \eqref{arealaw}), where $ A = 
4\pi \tilde r_A^2 $ is the area of the black hole horizon and $ G=\ell_p^2 $ is 
the gravitational constant. Hence, incorporating  this in a cosmological 
framework, the entropy of the apparent horizon is
\begin{equation}
\label{BHent}
    S_h=\frac{A}{4G}\,.
\end{equation}

Let us now apply the  first law of thermodynamics to the apparent horizon. As 
the Universe evolves over an infinitesimally small time interval $dt$, the heat 
flow through the horizon
is easily found to be \cite{Cai:2005ra}
\begin{equation}
\label{fltherm}
-\delta Q\,=\,dE\,=\,T_h dS\,+\, W dV\,, 
\end{equation}
where $dE$  represents the decrease in the total energy content of the Universe, 
and $dV$ is the change in its volume. The corresponding work density, which 
arises from the variation in the apparent horizon radius and, similar to fluid 
dynamics, effectively replaces the role of pressure, is given by
$\mathcal{W} = -\frac{1}{2} \text{Tr}(T^{\mu\nu}) = \frac{1}{2}  (\rho_m - 
p_m)$.  

If we define the total energy of  the Universe as $E=\rho_m V$, where 
$V=\frac{4}{3}\pi \tilde r_A^3$ is the volume, then 
differentiating yields  $dE=4\pi \tilde r_A^2 \rho_m d\tilde 
r_A+\frac{4\pi}{3}\tilde r_A^3\dot\rho_m dt$. Applying Eq. \eqref{cont}, this 
expression can be further manipulated to give
\begin{equation}
\label{Edif}
    dE\,=\,4\pi \tilde  r_A^2 \rho_m d\tilde r_A-4\pi H \tilde 
r_A^3\left(\rho_m+p_m\right)dt\,.
\end{equation}
On the other hand, the area law \eqref{BHent} gives the entropy variation 
\begin{equation}
\label{entdif}
    dS=\frac{2\pi}{G}\tilde r_A d\tilde r_A\,.
\end{equation} 
By substituting Eqs. \eqref{temp}, \eqref{Edif}, and \eqref{entdif} into Eq. 
\eqref{fltherm}, and performing some algebra, we obtain
\begin{equation}
\label{drdrho}
   \frac{d\tilde r_A}{\tilde r_A^3} \,=\,-\frac{4\pi G}{3} d\rho_m\,, 
\end{equation}
where, on the right-hand side,  we have substituted $dt$ back in terms of 
$d\rho_m$ using the continuity equation. Integration of Eq. \eqref{drdrho} 
finally leads to the first Friedmann equation
\begin{eqnarray}
\label{FFE}
    H^2-\frac{\Lambda}{3}\,=\,\frac{8\pi G}{3}\rho_m\,,
\end{eqnarray}
where the cosmological  constant $\Lambda$ arises as an integration constant. 
Differentiating with respect to $t$ leads to the second Friedmann equation
\begin{equation}
\label{SFE}
    \dot H\,=\,-4\pi G\left(\rho_m+p_m\right).
\end{equation}
In summary, as we saw, the Friedmann equations were extracted not by action 
variation, but by application of the first law of thermodynamics on the 
Universe apparent horizon.

\subsection{Modified Friedmann equations through 
generalized $S_{\delta,\epsilon}$ entropy}

In the previous subsection we presented the basics of the 
gravity-thermodynamics conjecture, showing that  the first law of black hole 
thermodynamics, when combined with the Bekenstein-Hawking formula for the 
entropy of the Universe's horizon, leads to the standard cosmological equations 
of $\Lambda$CDM paradigm. Hence, we deduce that   the cosmological equations  
would be modified if an alternative, non-holographic state-space scaling were to 
be considered. Toward this objective, in the following we will repeat the same 
procedure, retaining the form of the first law of 
thermodynamics given in Eq. \eqref{fltherm} and the definition of horizon 
temperature in Eq. \eqref{temp}, while employing the generalized entropy 
relation  \eqref{Sde}.\footnote{For completeness let us comment here 
that in the literature there is a discussion on whether one modified 
thermodynamic expression should be accompanied by modifications of all other 
thermodynamic quantities and/or by modifications of the fundamental 
thermodynamic laws, see e.g. \cite{Nojiri:2021czz}.}

The generalized entropy 
relation  \eqref{Sde} yields
\begin{equation}
    dS_{\delta,\epsilon}\,=\,  2\left[\left(4\pi\right)^\delta 
\gamma_\delta\hspace{0.2mm} \delta\hspace{0.2mm}\tilde r_A^{2\delta-1} \,+\, 
\left(4\pi\right)^\epsilon \gamma_\epsilon\hspace{0.2mm} 
\epsilon\hspace{0.2mm}\tilde r_A^{2\epsilon-1}
    \right]d\tilde r_A\,.
\end{equation}
 Inserting this relation into the first law of thermodynamics \eqref{fltherm} we 
finally acquire
\begin{equation}
\label{FeqSde}
    \frac{G}{\pi\tilde r_A^4}\left[\frac{\left(4\pi\right)^\delta\gamma_\delta\hspace{0.2mm} \delta\hspace{0.2mm}\tilde r_A^{2\delta}}{2-\delta}+\frac{\left(4\pi\right)^\epsilon\gamma_\epsilon\hspace{0.2mm} \epsilon\hspace{0.2mm}\tilde r_A^{2\epsilon}}{2-\epsilon}\right]\,=\,\frac{8\pi G}{3}\rho_m\,+\,C\,,
\end{equation}
where the integration constant $C$ has dimensions $[L^{-2}]$. By further defining 
\begin{equation}
    \alpha_{\delta}\,\equiv\, \frac{\left(4\pi\right)^{\delta} G\hspace{0.2mm}\gamma_\delta\hspace{0.2mm}\delta}{\left(2-\delta\right)\pi}
\end{equation}
and  similarly  $\alpha_{\epsilon}\equiv\alpha_{\delta\rightarrow\epsilon}$, 
Eq. \eqref{FeqSde} can be cast in the form
\begin{equation}
\label{FMFE}
    H^2\,=\,\frac{8\pi G}{3}\left(\rho_m+\rho_{DE}\right),
\end{equation}
where we have used  \eqref{ra}, and where we have defined an effective dark 
energy sector with energy density
\begin{equation}
    \rho_{DE}\,=\,  \frac{3}{8\pi G} \left[H^2\left(1-\alpha_\delta 
H^{2\left(1-\delta\right)}-\alpha_\epsilon 
H^{2\left(1-\epsilon\right)}\right)\,+\,C
    \right].
    \label{EfDEd}
\end{equation}

As we saw, the new generalized entropy expression through the 
gravity-thermodynamics approach,  lead to an effective dark 
energy of entropic origin. It should be noted that in the three limiting cases 
\eqref{limits} we obtain
$\rho_{DE}=\frac{3}{8\pi 
G}\hspace{0.3mm}C=const.$  which ensures 
consistency with the standard Friedmann equation \eqref{FFE}, provided that 
$C=\Lambda/3$. 

Differentiating Eq. \eqref{FMFE} yields the second modified Friedmann equation 
as
\begin{equation}
\label{SMFE}
    \dot H\,=\,-4\pi G\left(\rho_m+p_m+\rho_{DE}+p_{DE}\right),
\end{equation}
where we have introduced the effective entropic  dark energy pressure 
\begin{eqnarray}
\nonumber
    &&
\!\!\!\!\!\!\!\!\!\!\!    
    p_{DE}=-\frac{1}{8\pi G}\Big\{
    \Lambda + 3H^2\left[1-\alpha_\delta H^{2(1-\delta)}-\alpha_\epsilon  
H^{2(1-\epsilon)}\right]
    \\ 
    && \!\!\!\!\!\!\!\!\!  +\left.2\dot H \left[1-\alpha_\delta (2-\delta)  
H^{2(1-\delta)}-\alpha_\epsilon (2-\epsilon) H^{2(1-\epsilon)}
    \right]
    \right\}.
    \label{EfDEp}
\end{eqnarray}
As expected,  when any of the three conditions in Eq. \eqref{limits} hold  we 
acquire  $p_{DE}=-\frac{\Lambda}{8\pi G}$, thereby recovering the unmodified 
Friedmann equation \eqref{SFE}. 
Lastly, from Eqs. \eqref{EfDEd}, \eqref{EfDEp}  we can read   the effective 
equation
of state (EoS), namely
\begin{eqnarray}
\label{EoS}
   && \!\!\!\!\!\!\!\!\! w_{DE}\equiv\frac{p_{DE}}{\rho_{DE}}\\[2mm]
    \nonumber
   && \!\!\!\!\!\!\!\!\! =-1-\frac{2\dot H\left[1-\alpha_\delta (2-\delta) 
H^{2(1-\delta)}-\alpha_\epsilon (2-\epsilon)H^{2(1-\epsilon)}
    \right]}
    {\Lambda +3H^2\left(1-\alpha_\delta H^{2(1-\delta)}-\alpha_\epsilon H^{2(1-\epsilon)}\right) }\,,
\end{eqnarray}
which reproduces the cosmological constant-like behavior $w_{\Lambda}=-1$  
in the limiting cases \eqref{limits}.

\section{Cosmological Implications}
\label{Imp}

In the previous section we applied the gravity-thermodynamics conjecture with 
the generalized $S_{\delta,\epsilon}$ entropy   (\ref{genentropyexpr}), 
resulting to a modified cosmological scenarios, namely to the modified 
Friedmann equations   \eqref{FMFE} and \eqref{SMFE}, with an effective dark 
energy sector described by\eqref{EfDEd},\eqref{EfDEp}. These equations reduce 
to $\Lambda$CDM paradigm in the three limiting cases \eqref{limits}, in which 
case $S_{\delta,\epsilon}$ reverts to its conventional holographic scaling and 
recovers the standard Bekenstein-Hawking formula. Hence, in this section, we 
investigate the resulting cosmological dynamics and sd derive analytical 
solutions.

We consider   dust-like matter, namely we impose $p_m =0$. In this 
case Eq. \eqref{cont} yields the well-known scaling
$\rho_m = \frac{\rho_{m0}}{a^3}$,  where $\rho_{m0}$ denotes the matter energy 
density at present time and we have implicitly assumed $a_0=1$ (hereafter, the 
subscript ``0'' represents the value of a given quantity at present time). 
Additionally, we define the fractional density parameters 
\begin{eqnarray}
\label{Omm}
    \Omega_m&=&\frac{8\pi G}{3H^2}\hspace{0.2mm}\rho_m\,, \\[2mm]
     \Omega_{DE}&=&\frac{8\pi G}{3H^2}\hspace{0.2mm}\rho_{DE}\,,
     \label{Omd}
\end{eqnarray}
for   matter and dark energy sectors respectively. Hence, the first Friedmann 
equations gives just  $\Omega_m+\Omega_{DE}=1$. 
Noting that $\Omega_m=\Omega_{m0}H_0^2/(a^3 H^2)$, we can rewrite the Hubble rate in the form
\begin{equation}
\label{Hnew}
    H\,=\,H_0\sqrt{\frac{\Omega_{m0}}{a^3\left(1-\Omega_{DE}\right)}}=H_0\sqrt{\frac{\Omega_{m0}\left(1+z\right)^3}{\left(1-\Omega_{DE}\right)}}\,,
\end{equation}
where  in the last  step  we have used  \eqref{reds}. 

Plugging  
\eqref{EfDEd} into \eqref{Omd}, and using  \eqref{Hnew}, we finally acquire
\begin{eqnarray}
\label{Om2}
&&\!\!\!\!\!\! 1-\Omega_{DE}=-\frac{\Lambda\left(1-\Omega_{DE}\right)}{
3H_0^2 \Omega_{m0}\left(1\!+\!z\right)^3}+\alpha_\delta\left[\frac{H_0^2 
\Omega_{m0}\left(1\!+z\!\right)^3}{
1-\Omega_{DE}}\right]^{1-\delta}\nonumber \\ 
&&
\ \ \ \ \ \ \ \ \ \ \ \  \,
+\alpha_\epsilon\left[\frac{H_0^2 
 \Omega_{m0}\left(1\!+\!z\right)^3}{1-\Omega_{DE}}\right]^{1-\epsilon},
\end{eqnarray}
along with the condition
\begin{equation}
    \Lambda\,=\,3H_0^2\left(\alpha_\delta H_0^{2(1-\delta)}+\alpha_\epsilon H_0^{2(1-\epsilon)}-\Omega_{m0}\right)\,,
\end{equation}
which arises from  application of Eq. \eqref{Om2} at $z=0$. The above relation 
eliminates one of the free parameters of the model, as it allows the 
cosmological constant to be expressed in terms of the entropic exponents \( 
\delta , \epsilon \), and the observationally determined quantities \( H_0 
\) and \( \Omega_{m0} \). Hence, we have resulted to an analytical solution for 
the evolution of the dark energy density parameter $\Omega_{DE}$ as a function 
of the redshift.  In the following subsections we elaborate it in detail.

\begin{figure}[t]
    \centering
\includegraphics[width=7.cm]{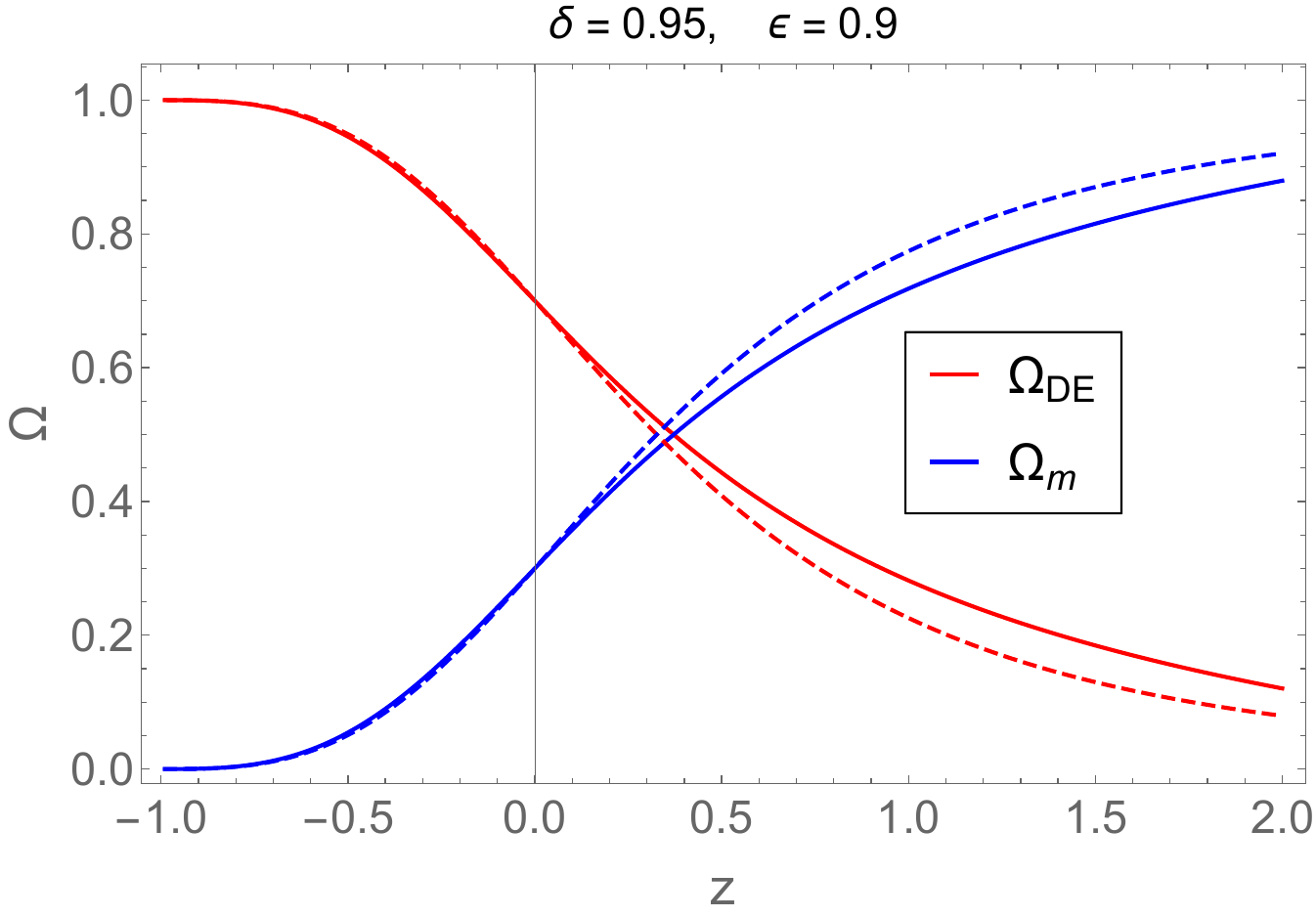}
\vspace{3mm}
\includegraphics[width=7.cm]{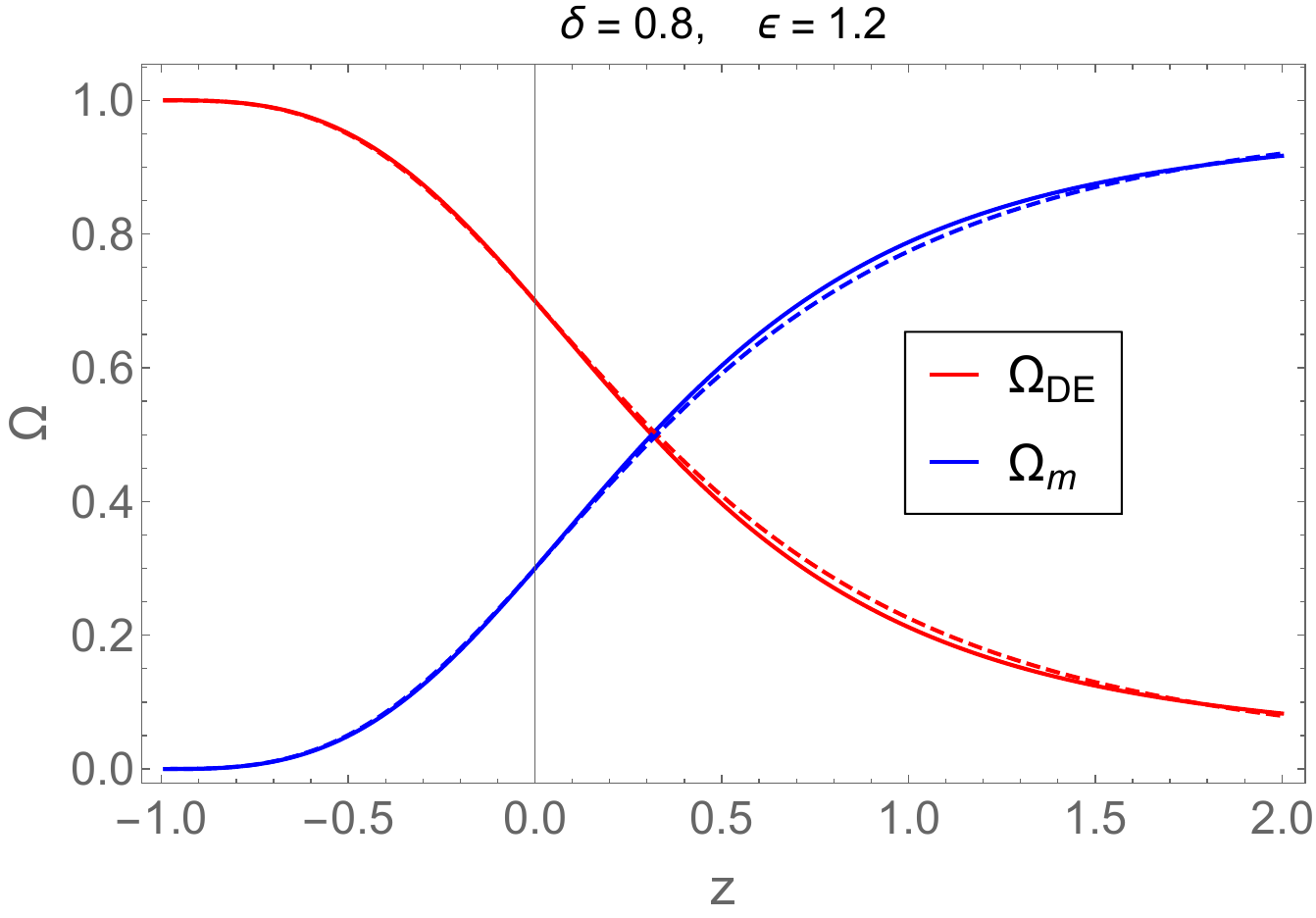}
\vspace{3mm}
\includegraphics[width=7.cm]{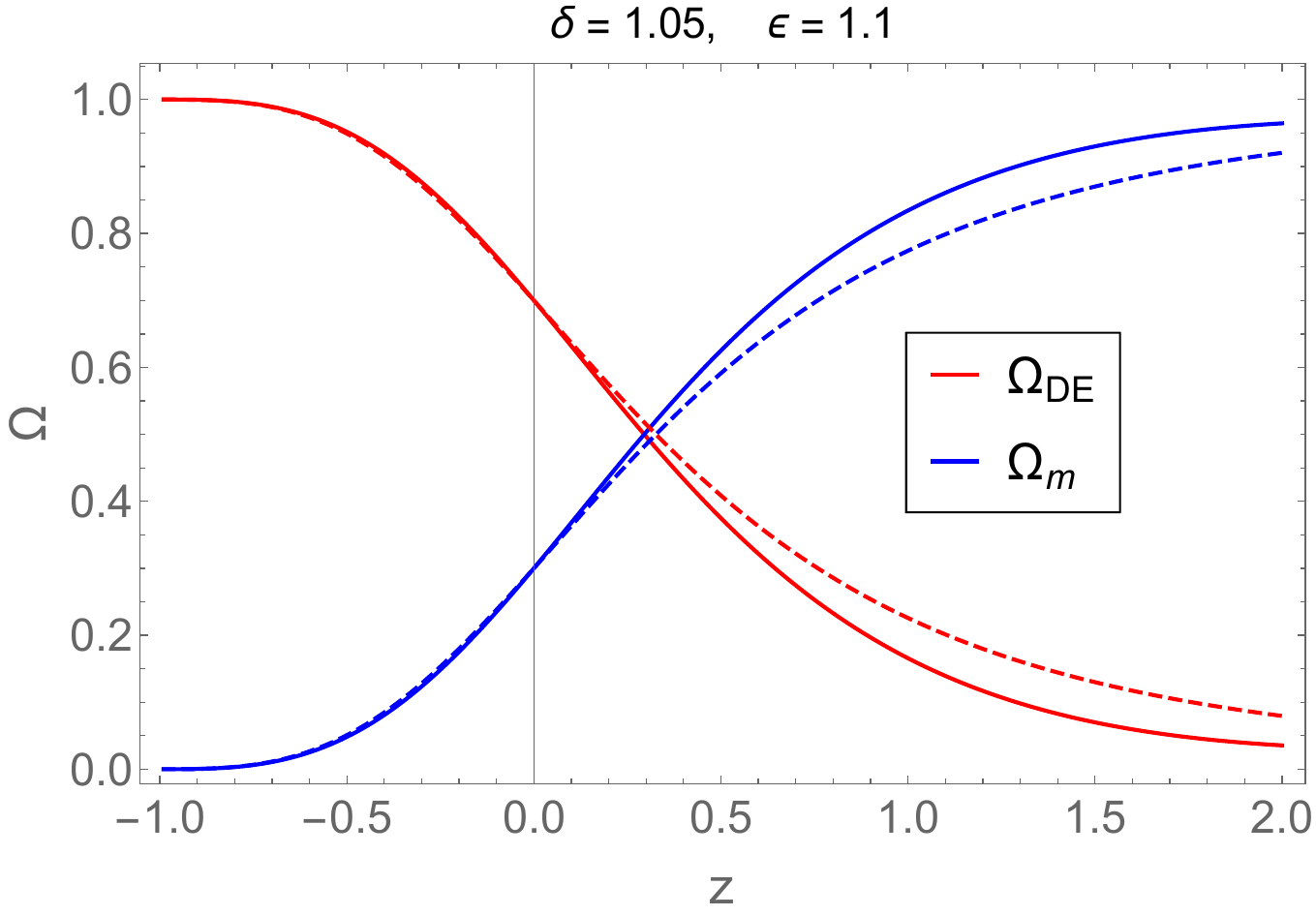}
\caption{\it{The evolution of the effective dark-energy density $\Omega_{DE}$ 
(solid red curve) and of the matter density parameter $\Omega_{m}$ (solid blue 
curve) as   functions of the redshift $z$. We set 
$\gamma_\delta=\left(1/4G\right)^\delta/2$, 
$\gamma_\epsilon=\gamma_{\delta\rightarrow\epsilon}$ and  $G=1$. Finally, the 
dashed curves represent the $\Lambda$CDM 
predictions. }}
    \label{Fig1}
\end{figure}

\subsection{Cosmological evolution in the general case}

Let us first study the general case. 
Since the $\Lambda$CDM paradigm is recovered for the parameter choices   given 
in   \eqref{limits}, our goal is to explore the role of the new entropic 
exponents on the Universe evolution. 

Since we have extracted the analytical  of the dark energy density parameter 
$\Omega_{DE}$ as a function 
of the redshift  in equation \eqref{Om2} above, in Fig. \ref{Fig1} 
we display  the behavior of $\Omega_{DE}$ (red lines) and 
$\Omega_m=1-\Omega_{DE}$ (blue 
lines) for different combinations of 
$\delta$ and $\epsilon$,   imposing the  intial condition $\Omega_{m0} \simeq 
0.3$. In the upper panel  we consider the case $\delta, \epsilon < 1$, 
corresponding to a total number of microstates describing the Universe being 
lower than in the standard holographic-like scenario. As observed, the effective 
dark energy density decreases more slowly than the $\Lambda$CDM prediction 
(dashed red line) at high redshift.
The opposite behavior  occurs in the lower panel, where $\delta, \epsilon > 1$. 
The intermediate regime, with $\delta < 1$ and $\epsilon > 1$, is explored in 
the middle panel,  which reveals that the two entropic corrections nearly cancel 
each other, resulting to an evolution closely resembling $\Lambda$CDM. 
Additionally, independently of the parameter values, it can be observed that 
$\Omega_{DE} \to 1$ and $\Omega'_{DE} \to 0$ (prime marks $z$-derivative) as $z 
\to -1$, suggesting a scenario in which DE completely dominates the energy 
content of the Universe in the far future. This analysis underscores the 
nontrivial impact of entropic parameters on the evolution of the DE sector in 
the Universe.

We now focus on the important  observable quantity, namely the dark energy 
equation-of-state parameter 
$w_{DE}$ given in  \eqref{EoS}.  Differentiating   \eqref{Hnew} with 
respect to $t$, we are led to
\begin{equation}
    \dot H\,=\,-\frac{H^2}{2\left(1-\Omega_{DE}\right)}\left[3\left(1-\Omega_{DE}\right)+\left(1+z\right)\Omega'_{DE}\right],
    \label{Hdot}
\end{equation}
where we have used that  $\dot \Omega_{DE}=-\Omega'_{DE} H (1+z)$. Combining 
  \eqref{EoS}, \eqref{Hnew} and \eqref{Hdot}, we  acquire
\begin{eqnarray}
\label{omde2}
&&
\!\!\!\!\!\!\!\!\!\!\!\!\!\!\!\!\!\!\!\!\!\!\!\!\!\!\!
w_{DE}=-1+\frac{h_m\left[3\left(1-\Omega_{DE}\right)+\left(1+z\right)\Omega'_{
DE}\right]}{\left(1-\Omega_{DE}\right)}   \nonumber\\[2mm]
&&
\ \ \ \  
\times\,\frac{\left[
1-\left(2-\delta\right)f_\delta-\left(2-\epsilon\right)f_\epsilon
    \right]}{\Lambda\left(1-\Omega_{DE}\right)+3h_m\left[1-f_{\delta}-f_{\epsilon}
    \right]}\,,
\end{eqnarray}
where we have introduced the shorthand notations
\begin{eqnarray}
h_{m}&\equiv&H_0^2\hspace{0.2mm}\Omega_{m0}\left(1+z\right)^3\,,\\[2mm]
\label{fd}
f_{\delta}&\equiv& \alpha_\delta \left(\frac{h_m}{1-\Omega_{DE}}\right)^{1-\delta}\,,\\[2mm]
\label{fe}
f_{\epsilon}&\equiv& f_{\delta\rightarrow\epsilon}\,.
\end{eqnarray}
It can be easily verified that  if any of the three conditions in  
\eqref{limits} is satisfied, then the extra terms in \eqref{omde2} vanishes 
and we 
recover $w_{DE} = -1 = \emph{const.}$, consistently with the $\Lambda$CDM 
scenario. Nevertheless, as the entropy deviates from the holographic scaling, 
$w_{DE}$ acquires a dynamical nature, exhibiting a much richer behavior compared 
to standard cosmology.

\begin{figure}[ht!]
    \centering
\includegraphics[width=8cm]{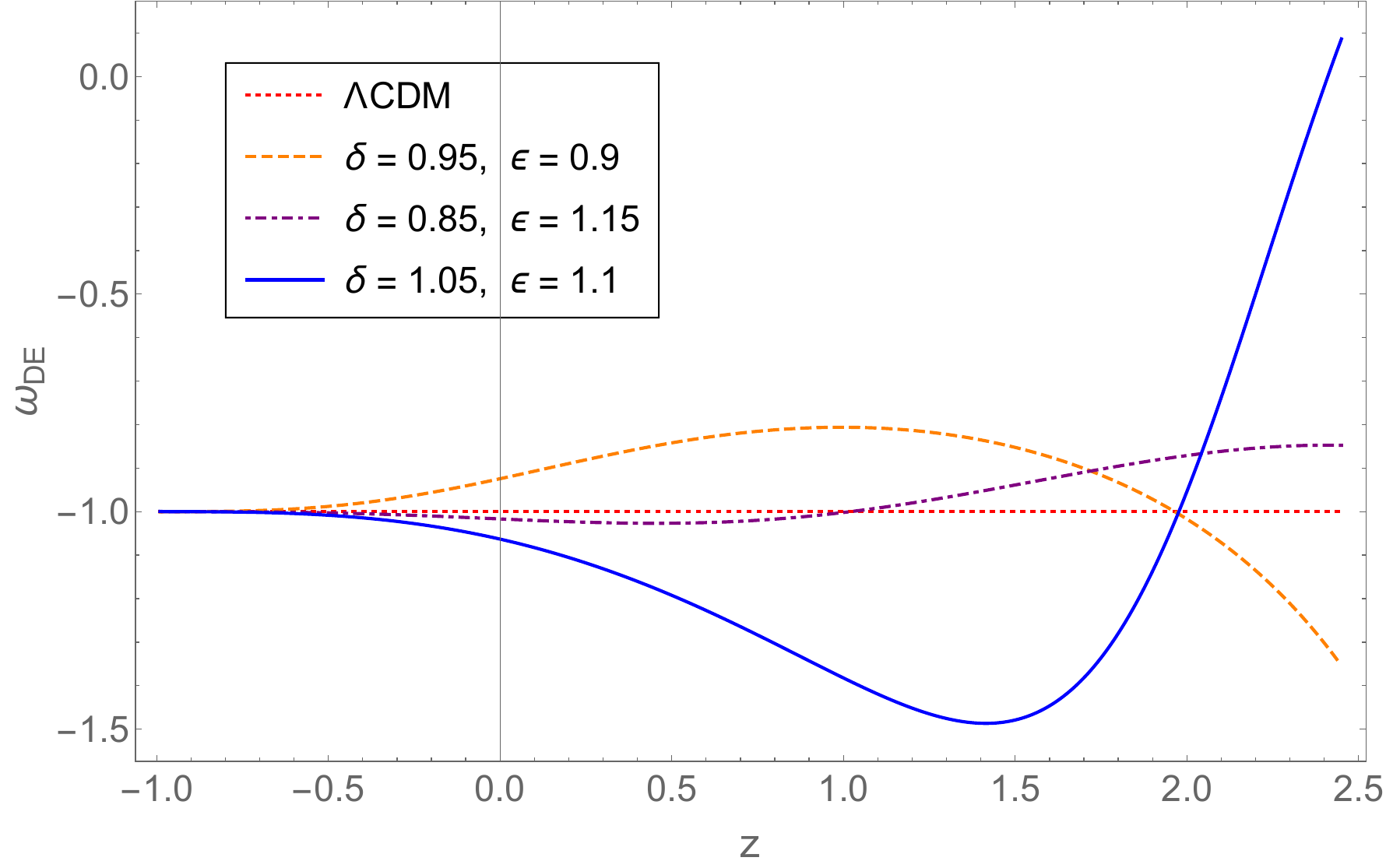}
\caption{\it{The evolution of       the dark-energy equation-of-state 
$w_{DE}$ as a function of the redshift $z$, for 
different combinations of $\delta$ and $\epsilon$. We set 
$\gamma_\delta=\left(1/4G\right)^\delta/2$, 
$\gamma_\epsilon=\gamma_{\delta\rightarrow\epsilon}$ and  $G=1$.}}
\label{Fig2}
\end{figure}
In order to investigate the impact of $S_{\delta,\epsilon}$ on $w_{DE}$,
in Fig. \ref{Fig2} we present its evolution   for various values of 
$\delta, \epsilon$.  As $z$ decreases, we observe that $w_{DE}(z)$  evolves 
from a quintessence-like behavior ($w_{DE} > -1$) to a phantom-like 
behavior ($w_{DE} < -1$) for $\delta, \epsilon > 1$ (solid blue curve).
The fact that dark energy can exhibit a phantom-like EoS at present time 
represents a significant advantage of this model, as it would help to alleviate 
the $H_0$  tension between low and high redshift observations 
\cite{DiValentino:2020naf}. 
 Indeed, phantom  dark energy is one of the mechanisms that  can amplify 
late-time cosmic acceleration, thereby increasing the value of $H_0$ favored by 
CMB data and releasing the tension with local measurements of the Hubble rate 
\cite{Abdalla:2022yfr}.

The reverse transition, namely from phantom to quintessence 
domain, occurs when $\delta, \epsilon < -1$ (dashed orange curve).  
Finally, in the intermediate regime where $\delta < 1$ and $\epsilon > 1$ 
(dot-dashed purple curve), the nature of the entropic dark energy is 
approximately equal to that of the cosmological constant  (i.e, $w_{DE} \approx 
w_\Lambda = 
-1$), consistently with the previously discussed behavior of $\Omega_{DE}$.
At present time, for the given parameter values, we find that $w_{DE}$ lies 
within the range $[-1.06, -0.92]$, to be compared with $w_{DE}\sim -1.03$ 
suggested by multiple independent experimental probes \cite{Escamilla:2023oce}.

Lastly, taking the limit $z\rightarrow-1$,  we see that 
in each of the aforementioned scenarios we have  $w_{DE}\rightarrow-1$ . This 
implies that, although our model 
deviates from $\Lambda$CDM at intermediate times, the effective dark energy 
density will ultimately stabilize at the cosmological constant value in the 
asymptotic future, driving the Universe towards a de Sitter  phase, 
regardless of the entropic exponents. The fact that a $\Lambda$-dominated 
Universe acts as a stable late-time attractor is another advantageous feature of 
the scenario at hand, 
ensuring a smooth cosmic evolution and avoiding any catastrophic fate (such as 
Big Rip).

We mention here that the present scenario exhibits a richer phenomenology 
compared to other recent cosmological models based on extended entropies. 
For instance, a comparison with \cite{Ebrahimi:2024zrk} reveals that, for 
realistically small deviations from holographic scaling, the Ghost Dark Energy  
in pure Tsallis cosmology appears to exhibit phantom-like behavior only in the 
far future, even in the presence of interactions between the dark sectors.  On 
the other hand, within the framework of Barrow entropy-based cosmology, the 
restriction of the parameter space to the case $\Delta>0$ reproduces only the 
first of the three scenarios described above in our model (namely for 
$\delta,\epsilon>1$) \cite{Saridakis:2020lrg}. 

\begin{figure}[t]
    \centering
\includegraphics[width=7.5cm]{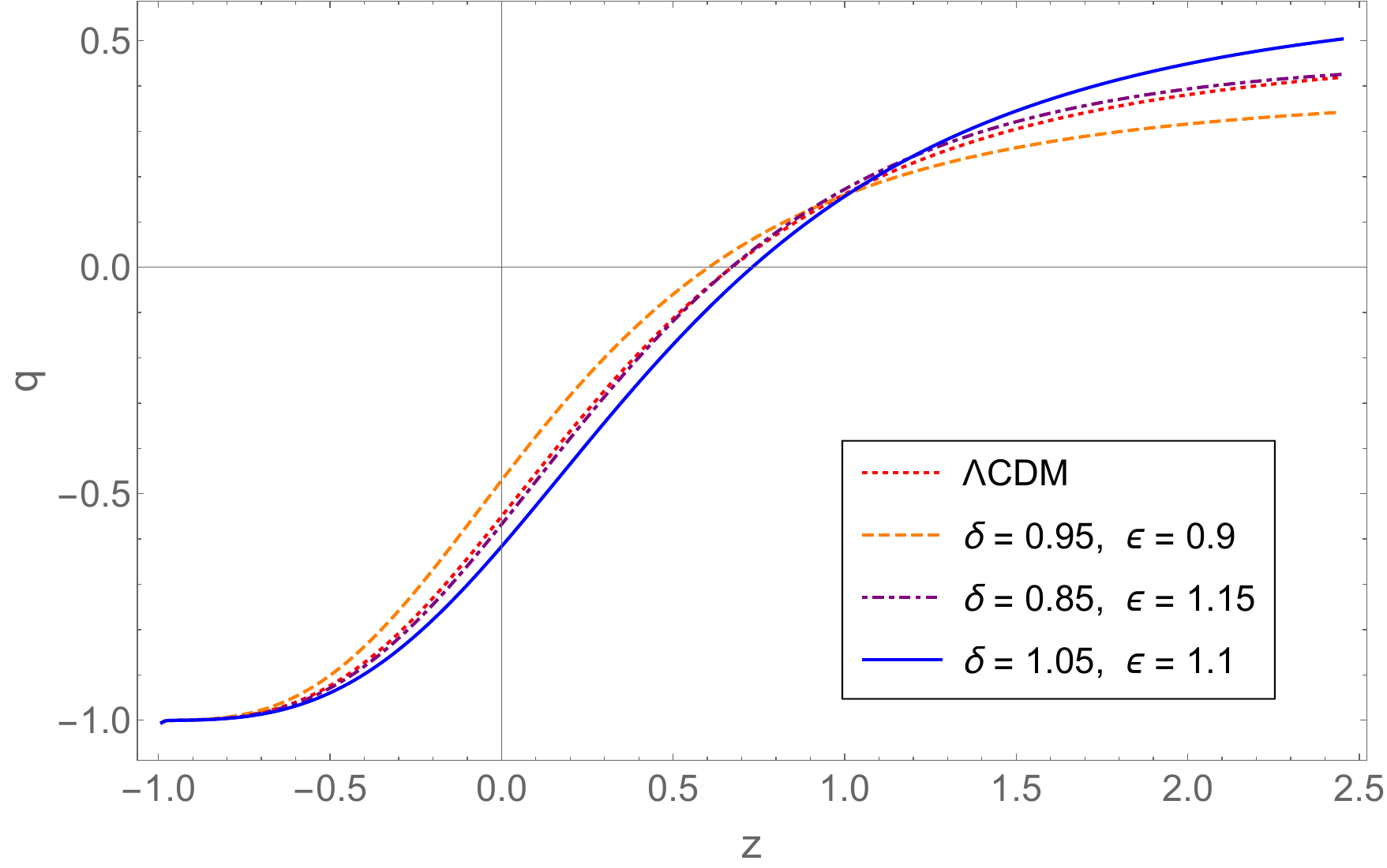}
\caption{\it{The evolution of the deceleration parameter $q$ as a function of 
the  redshift $z$, for different combinations of $\delta$ and $\epsilon$. We 
set $\gamma_\delta=\left(1/4G\right)^\delta/2$, 
$\gamma_\epsilon=\gamma_{\delta\rightarrow\epsilon}$ and  $G=1$.}}
\label{dec}
\end{figure}

We proceed by studying the evolution of the deceleration 
parameter 
\begin{equation}
\label{decpar}
    q\equiv-\frac{\ddot a}{a H^2}
    =-1-\frac{\dot H}{H^2}\,,
\end{equation} 
which provides  information on the Universe  expansion rate, indicating whether 
it is slowing down ($q_0 > 0$), accelerating ($q_0 < 0$) or proceeding at a 
constant rate ($q_0 = 0$). Using Eqs. \eqref{Hnew} and \eqref{Hdot}, we find 
that 
\begin{equation}
    q\,=\,-1+\frac{1}{2\left(1-\Omega_{DE}\right)}\left[3 
\left(1-\Omega_{DE}\right)+\left(1+z\right)\Omega'_{DE}
    \right].
\end{equation}
The evolution of \( q \) as a function of \( z \)  is shown in Fig.~\ref{dec} 
for different values of the model parameters. Notably, the present value of the 
deceleration parameter is found to be within the range $[-0.61, -0.47]$, which 
aligns with both the $\Lambda$CDM prediction ($q_0\approx-0.55$ 
\cite{Camarena:2019moy}) and the recent estimate from the combined analysis of 
Hz, SNe and BAO data ($q_0=-0.54^{+0.21}_{-0.21}$, see \cite{Koussour:2024kxd} 
and references therein). In addition, the transition redshift $z_{tr}$ can take 
values within the range $[0.59,0.73]$, which is likewise in agreement with the 
combined Hz + SNe + BAO datasets ($z_{tr}=0.67^{+0.08}_{-0.07}$).
Remarkably, these findings are also consistent with previous results  in the 
literature \cite{Cunha:2008ja,Camarena:2019moy,Mehrabi:2021cob}, reinforcing our 
understanding of the Universe expansion dynamics.
At this point we mention that, in all the aforementioned examples, we have kept 
the values of $\gamma_\delta$ and $\gamma_\epsilon$ fixed. Clearly, allowing 
such parameters to vary would enhance the  capability of the scenario and 
enrich the predicted cosmological behaviors.

For completeness, to further assess the consistency of the  scenario with 
observed phenomenology and improve its theoretical robustness, we   confront it 
with Supernovae type Ia (SN Ia) data. In these observational dataset the 
apparent luminosity $l(z)$ or, equivalently, the apparent magnitude $m(z)$ is 
measured as a function of redshift. These quantities are directly linked to the 
luminosity distance through the relation
\begin{eqnarray}
  && 
  \!\!\!\!\!\!\!\!\!\!\!\!\!\!\!\!\!\!\!\!\!\!\!\!\!  
  2.5\,\mathrm{Log}\left[\frac{L}{l(z)}\right]=\mu(z) \equiv 
m(z)-M\nonumber\\
&&
\ \ \ \ \ \ \ 
=5\mathrm{Log}\left[\frac{d_L(z)_{obs}}{Mpc}\right]+25,
\end{eqnarray}
where $M$ and $L$ denote the absolute magnitude and luminosity,  respectively. 
Furthermore, for any theoretical model  the predicted luminosity distance 
$d_L(z)_{th}$ can be determined based on the expected evolution of the Hubble 
function as
\begin{equation}
d_L(z)_{th}\,\equiv\,\left(1+z\right)\int_0^z\,\frac{dz'}{H(z')}\,.
\end{equation}

\begin{figure}[t]
    \centering
\includegraphics[width=8cm]{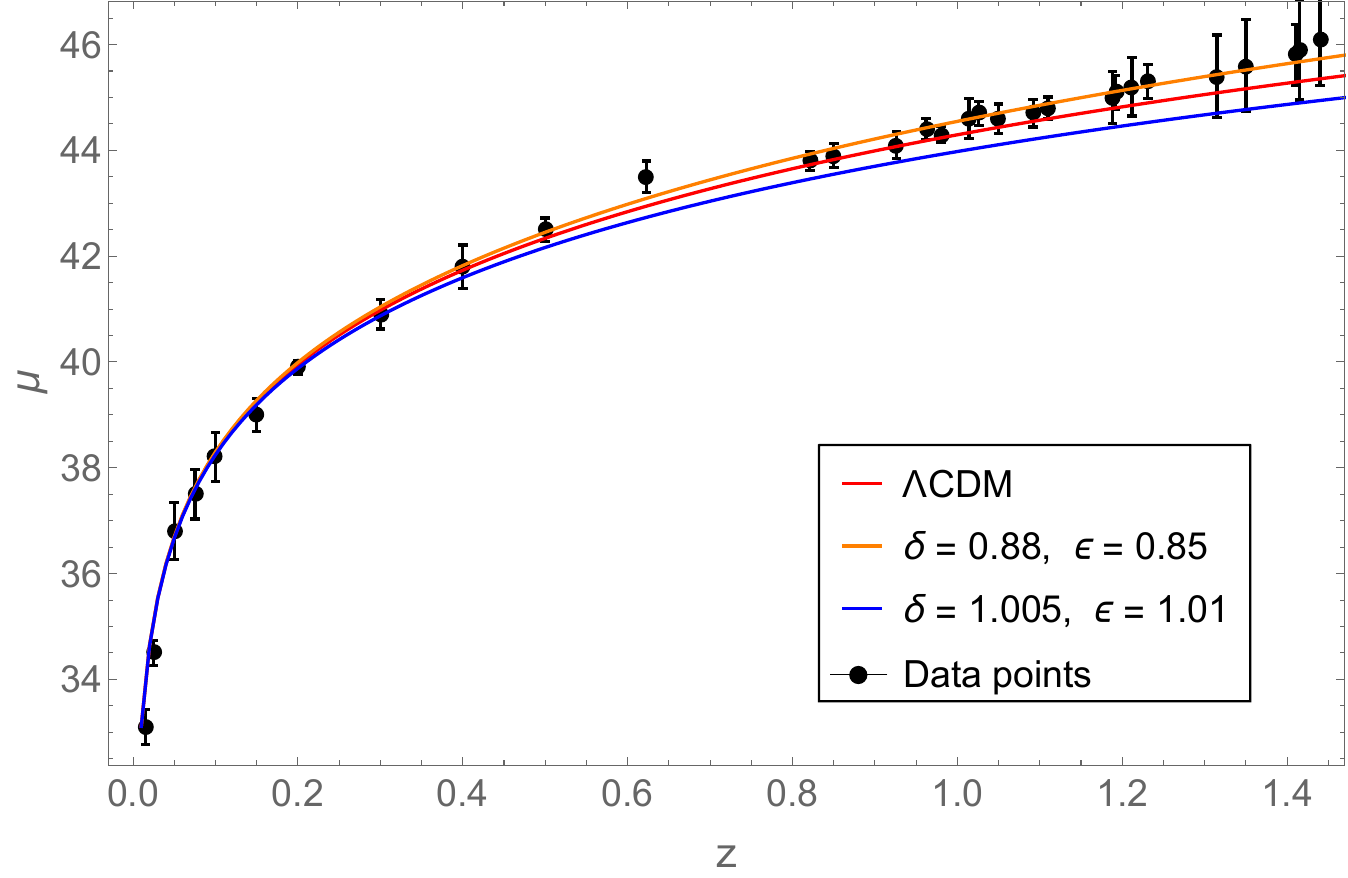}
\caption{\it{The theoretically  predicted distance modulus $\mu$ as a function 
of the redshift $z$, for different combinations of $\delta$ and $\epsilon$. The 
dots represent a set of observational SN Ia data points from 
\cite{Amanullah:2010vv, SupernovaCosmologyProject:2011ycw}. We set $\gamma_\delta 
= \left(\frac{1}{4G}\right)^\delta/2$, $\gamma_\epsilon = \gamma_{\delta \to 
\epsilon}$, and $G = 1$.}}
\label{dismod}
\end{figure}

Within our framework,  $d_L(z)_{th}$ can be easily calculated   using  
\eqref{Hnew} and \eqref{Om2}. In Fig.~\ref{dismod} we present the theoretical 
distance modulus $\mu(z)$ for different values of the model parameters, along 
with the corresponding $\Lambda$CDM prediction, overlaid with SN Ia 
observational data points extracted from 
\cite{Amanullah:2010vv,SupernovaCosmologyProject:2011ycw}. As observed, for 
appropriate values of $\delta, \epsilon < 1$ (orange curve), the agreement with 
the experimental data is slightly better than that of the $\Lambda$CDM 
prediction (red curve), particularly at high redshifts.  Nevertheless, in order 
to extract a complete result, a full observational confrontation and a 
statistical analysis is needed, alongside the application of various 
information criteria, such as the  Akaike Information Criterion (AIC)
 and the Bayesian Information Criterion (BIC).

\subsection{Cosmological evolution in the case $\Lambda=0$}

In the previous subsection we investigated the general case  in which 
the Friedmann equations are modified within the framework of nonextensive 
thermodynamics. In particular, we considered the case where the cosmological 
constant is explicitly present, arising as an integration constant, i.e. 
  we   had a scenario that encompasses $\Lambda$CDM cosmology as a special 
case. In this subsection  we   focus on a more radical application of the 
proposed scenario, by assuming the absence of an explicit cosmological 
constant. Although this approach does not recover $\Lambda$CDM as a limiting 
case, we demonstrate that it nonetheless offers a framework where the model 
parameters $\delta$ and $\epsilon$ can be appropriately tuned to mimic its 
effects, maintaining consistency with observational data.

As a first step, we observe that in the case $\Lambda=0$, Eqs. \eqref{EfDEd} 
and \eqref{EfDEp} can be simplified to
\begin{eqnarray}
  \rho_{DE}&=& \frac{3}{8\pi G} \left[H^2\left(1-\alpha_\delta H^{2\left(1-\delta\right)}-\alpha_\epsilon H^{2\left(1-\epsilon\right)}\right)
    \right],\\[4mm]
       p_{DE}&=&-\frac{1}{8\pi G}\Big\{3H^2\left[1-\alpha_\delta H^{2(1-\delta)}-\alpha_\epsilon H^{2(1-\epsilon)}\right]
    \\[2mm]
    \nonumber
    &&\hspace{-3mm}+\left.2\dot H \left[1-\alpha_\delta (2-\delta) H^{2(1-\delta)}-\alpha_\epsilon (2-\epsilon) H^{2(1-\epsilon)}
    \right]
    \right\}.
\end{eqnarray}
Additionally, application of    Eq. \eqref{Om2} at $z=0$, now gives the 
constraint
\begin{equation}
\label{Lam0}
    \Omega_{m0}\,=\,\alpha_\delta H_0^{2(1-\delta)}+\alpha_\epsilon H_0^{2(1-\epsilon)}\,,
\end{equation}
which enables us to express one of the two constants, e.g., $\alpha_\delta$,  in 
terms of $\alpha_\epsilon$, $H_0$ and $\Omega_{m,0}$. 
Moreover, the effective dark-energy equation-of-state parameter \eqref{EoS} 
becomes
{\small{
\begin{equation}
    w_{DE}=-1-\frac{2\dot H\left[1-\alpha_\delta (2-\delta) 
H^{2(1-\delta)}-\alpha_\epsilon (2-\epsilon)H^{2(1-\epsilon)}
    \right]}
    {3H^2\left(1-\alpha_\delta H^{2(1-\delta)}-\alpha_\epsilon H^{2(1-\epsilon)}\right) }\,,
\end{equation}}}
while Eq. \eqref{Om2} for the dark energy density takes the form
\begin{equation}
1-\Omega_{DE}= 
f_\delta\,+\,f_\epsilon\,,
\end{equation}
with $f_\delta, f_\epsilon$   defined in   \eqref{fd} and \eqref{fe}, 
respectively. 

\begin{figure}[!]
    \centering
    \includegraphics[width=0.8\linewidth]{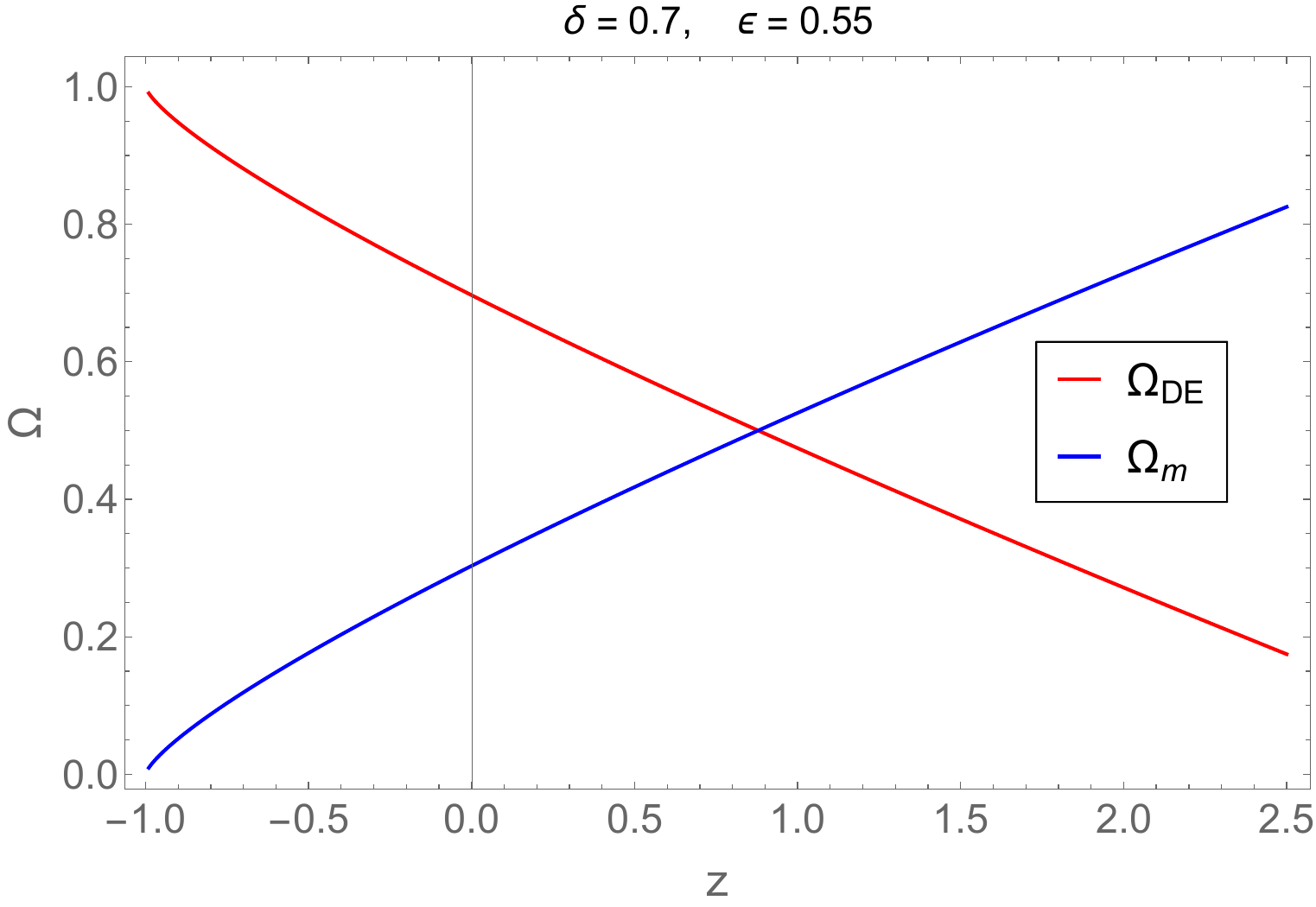}
\vspace{3mm}
\includegraphics[width=0.8\linewidth]{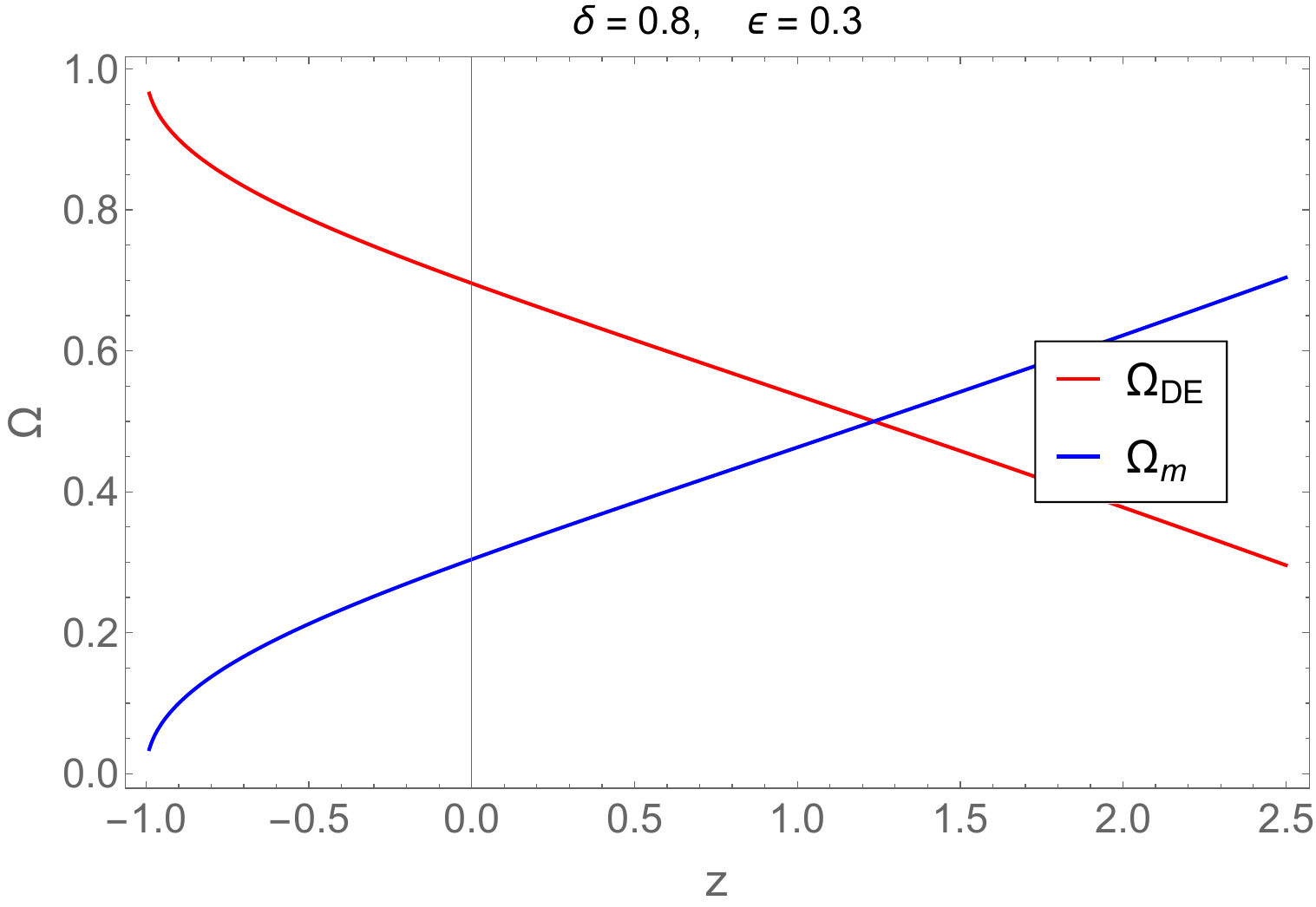}
        \caption{\it{The  evolution of the dark-energy density 
$\Omega_{DE}$ (solid red curve) and of the matter density parameter 
$\Omega_{m}$ (solid blue curve) as a function of the redshift $z$, in the 
specific case where  $\Lambda=0$. The entropic exponents are set according to 
Eq. \eqref{Lam0} to ensure $\Omega_{m0}\approx0.3$.}
}
    \label{La0}
\end{figure}
We emphasize that, in this case,  \(\Lambda\)CDM cosmology cannot be recovered 
for any choice of parameters, not even in the limiting cases given by 
Eq.~\eqref{limits} in which our extended entropy recovers the standard 
Bekenstein-Hawking one. However, interestingly enough, the usual thermal 
history of the Universe is preserved in this scenario. In particular, as 
depicted in Fig.~\ref{La0}, we acquire the expected sequence of matter and dark 
energy eras, with the Universe gradually approaching complete dark-energy 
domination in the far future (\(z \to -1\)). Notice that, in this analysis, we 
have considered only the case where \(\delta, \epsilon < 1\). However, similar 
arguments can be readily extended to other parameter combinations, provided that 
the condition \eqref{Lam0} remains satisfied. Furthermore, we mention that at 
high redshifts the evolution of \(\Omega_{DE}(z)\) may present issues, as this 
formulation either predicts the presence of early-time dark energy or leads to 
the unphysical scenario where \(\Omega_{DE}(z)\) becomes negative. However, 
these issues are naturally resolved when the radiation component is included, as 
it effectively regulates the early-time dynamics and ensures a cosmic evolution 
consistent with observations across all redshifts (see, for example, 
\cite{Lymperis:2018iuz}).

\begin{figure}[!]
    \centering
    \includegraphics[width=0.85\linewidth]{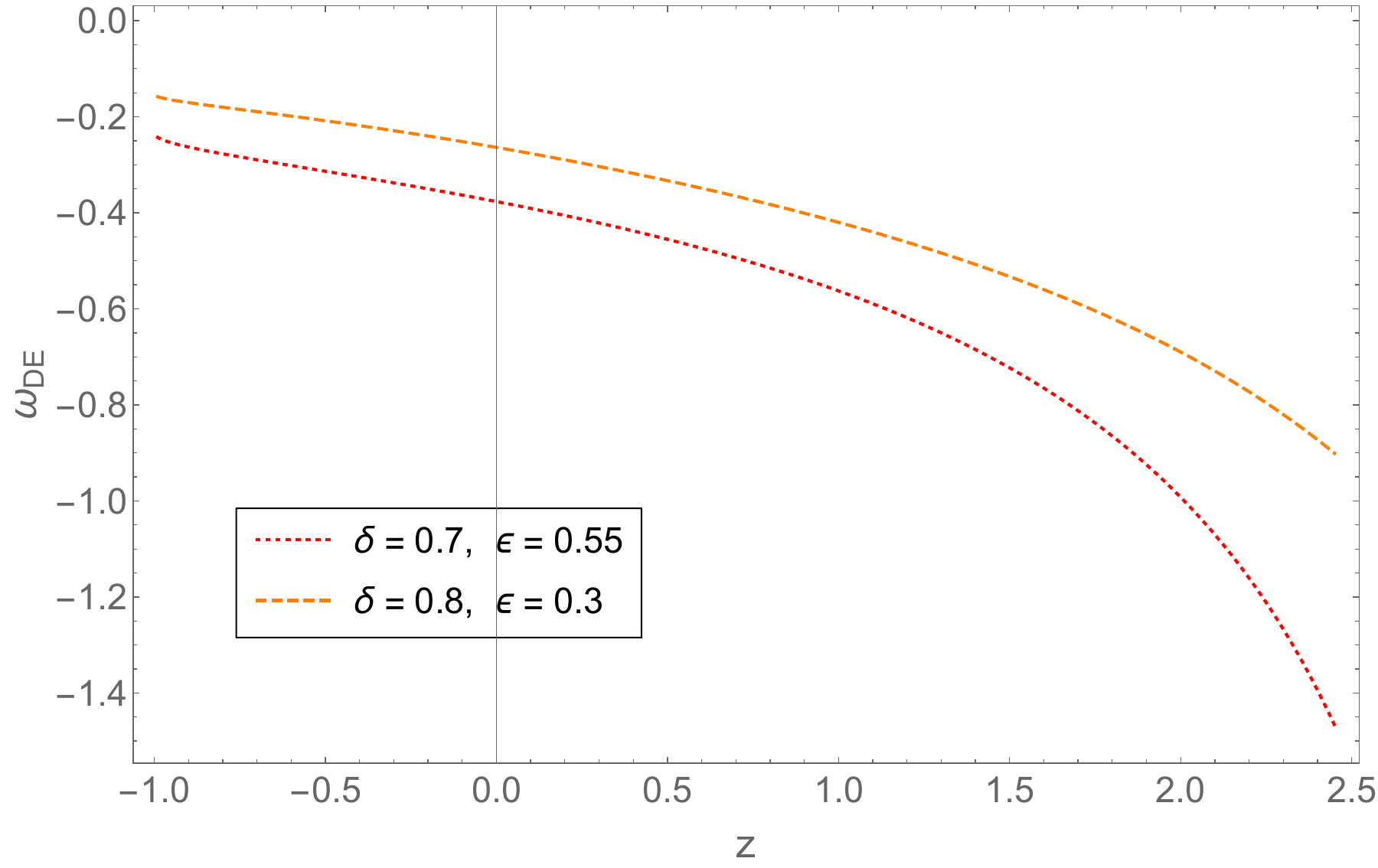}
    \caption{\it{The evolution of the dark-energy equation-of-state parameter 
$w_{DE}$ as a function of 
the redshift $z$ and for $\Lambda=0$. The entropic exponents are set according 
to Eq. \eqref{Lam0} to ensure $\Omega_{m0}\approx0.3$. Here, we have considered 
only the case where \(\delta,\epsilon< 1\).}
}
    \label{omLa0}
\end{figure}

Finally, Fig.~\ref{omLa0} shows that the asymptotic value of \(w_{DE}\) in 
the far future does not necessarily correspond to the cosmological constant 
value of \(-1\), indicating that the Universe does not ultimately evolve into a 
de Sitter space.

\section{Conclusions and Outlook}
\label{CandO}

A long-standing conjecture suggests  a fundamental connection between gravity 
and thermodynamics. In cosmology, this implies that the Friedmann equations may 
emerge from the first law of thermodynamics applied to the apparent horizon of 
the Universe. In this work, we   applied the general theory of entropy
in terms of the probability distribution over the accessible microstates, and 
by imposing violation of the separability requirement (expected in complex and 
interacting systems) and thus  considering a 
generalized microstate scaling, we resulted to a generalized entropy 
expression $S_{\delta,\epsilon}$.
Hence, by 
applying it in systems with boundaries, we extracted a generalized 
holographic-like area-law scaling   with two exponents.

We continued by incorporating  it within the gravity-thermodynamics framework in the 
Universe apparent horizon, and we  resulted to  a modified cosmological 
scenario. In particular, we obtained modified Friedmann equations 
 with additional terms, characterized by the 
two exponents $\delta$ and $\epsilon$, which eventually give rise to  an 
effective dark energy sector, and to a cosmological phenomenology richer than 
that of $\Lambda$CDM paradigm. 
Finally, as we showed, in the three limiting 
cases where the new generalized entropy reduces to the usual
Bekenstein-Hawking one, the  effective dark energy coincides with the 
cosmological constant and the $\Lambda$CDM paradigm is restored.

We extracted analytical expressions for the dark energy density parameter, as 
well as for the dark-energy equation-of-state parameter, as functions of the 
redshift. As we showed, the Universe experiences the usual thermal history, with 
the sequence of matter and dark-energy eras. Additionally, the dark energy can 
be quintessence-like, phantom-like or experience the phantom-divide crossing 
during its evolution, ultimately stabilizing at the cosmological constant value 
of $-1$ in the asymptotic far future. Interestingly, for suitable values of the 
model parameters, dark-energy exhibits a phantom-like EoS at present time, 
offering a potential mechanism to alleviate the $H_0$ tension. Furthermore, 
calculating the luminosity distance and comparing it with SN Ia data we offered 
an additional indication  that  the scenario is in agreement with observations.

Finally, for completeness, we explored also the more radical subcase in 
which the explicit integration constant, which play the role of a 
cosmological constant, is absent, and thus the effective dark-energy sector is 
constituted solely by the new terms. Although this case does not recover 
$\Lambda$CDM paradigm as a limiting scenario, for any choice of  the model 
parameters, it remains consistent with observations. Specifically, we saw 
that we can have the expected sequence of matter and dark-energy dominated 
epochs, before ultimately resulting  to a complete dark-energy domination in 
the far future. 

Several aspects  remain to be investigated. First, it would be valuable to 
conduct a comprehensive observational analysis using data from Cosmic Microwave 
Background (CMB) shift parameters, Baryon Acoustic Oscillations (BAO) and Hubble 
parameter observations, and extract the corresponding contour plots. This 
would allow for more precise constraints on the entropic exponents, while it 
would allow for a more accurate comparison with the $\Lambda$CDM paradigm, 
through the application of  various information criteria, such as the  
AIC and BIC ones. 
Furthermore, our model could be extended by incorporating   radiation, 
providing a more comprehensive understanding of the cosmological dynamics of the 
early Universe. From a more theoretical perspective, it would be interesting  
to investigate whether the corresponding gravitational model aligns with any 
extended theories already established in the literature through different 
frameworks. Work in these directions is presently under active investigation and 
will be presented elsewhere.

\section*{Acknowledgements}
The authors would like to thank Petr Jizba and Jan Korbel for insightful discussions.
The research of GGL is supported by the postdoctoral funding program of the University of Lleida.
The authors  acknowledge  the contribution  of 
the LISA CosWG, and of   COST 
Actions   CA21136 ``Addressing observational tensions in cosmology with 
systematics and fundamental physics (CosmoVerse)'', CA21106 ``COSMIC WISPers 
in the Dark Universe: Theory, astrophysics and experiments'', 
and CA23130 ``Bridging high and low energies in search of quantum gravity 
(BridgeQG)''.

\bibliography{Bib}

\end{document}